\documentclass[openacc]{rstransa}

\newcommand{\calo}{{\cal O}}

\begin{document}

\title{Black holes in the quantum universe}

\author{
Steven B. Giddings 
}

\address{Department of Physics, University of California, Santa Barbara, CA 93106, and\\
CERN, Theory Department,
1 Esplande des Particules, Geneva 23, CH-1211, Switzerland}

\subject{Quantum gravity, Black holes}


\corres{Steven B Giddings\\
\email{giddings@ucsb.edu}}

\begin{abstract}
A succinct summary is given of the problem of reconciling observation of black hole-like objects with quantum mechanics.
If quantum black holes behave like  subsystems, and also decay, their information must be transferred to their environments.  Interactions that accomplish this with `minimal' departure from a standard description are parameterized.  Possible sensitivity of 
gravitational wave or very long baseline interferometric observations to these interactions is briefly outlined.

\end{abstract}


\begin{fmtext}
\section{Introduction}

While there is spectacular evidence for objects that look and act very much like black holes in the Universe, there is no known description of their evolution that is consistent with the principles of quantum mechanics, which are believed to govern all physical law.  This appears to present a deep crisis in fundamental physics.

Black holes (BHs) certainly appear to exist.  Evidence includes jets shot from the centers of galaxies, like the 5000 lightyear long jet from the center of M87, thought to be generated by a supermassive black hole.  Stellar orbits have been observed about a highly compact object with approximate mass $4\times 10^6 M_\odot$ at the center of our own galaxy, and more recently an orbiting hotspot has been observed at a radius just a few times the corresponding Schwarzschild radius\cite{GRAVITY}.  Gravitational waves have now been observed by LIGO from what appear to be BH collisions, beginning with \cite{LIGOone}, and continuing with many more.  And remarkably, since this meeting, the Event Horizon Telescope has imaged the central object in M87, revealing an apparent "shadow" of a $6.5\times 10^9 M_\odot$ black hole\cite{EHT}.

To the best of our knowledge, all physical law needs to respect the principles of quantum mechanics, which are increasingly well-tested.  But, our present "standard" description of BHs, based on the geometrical spacetime

\end{fmtext}

\maketitle

\noindent
 picture of general relativity (GR), together with quantized fluctuations of other fields (including the gravitational field) on a background geometry, indicate a breakdown of quantum mechanics.  So, BHs appear to directly point to a deep inconsistency in our present quantum mechanical laws.

Specifically, all other observed phenomena in the Universe appear to be well-described within the framework of local quantum field theory (LQFT), on a semiclassical background geometry.  As a theoretical structure, LQFT arises from a set of basic principles:  the principles of quantum mechanics, the principles of relativity, and the principle of locality.  In particular, the latter two are directly associated with the statement that we work with classical geometry and topology.  BHs appear to reveal that this framework is inconsistent, and thus that these principles are inconsistent, and one or more of them must be modified.  We first turn to a lightening review of this crisis in physics.

\section{The unitarity crisis}\label{sec2}

\begin{figure}[!h]
\centering\includegraphics[width=6in]{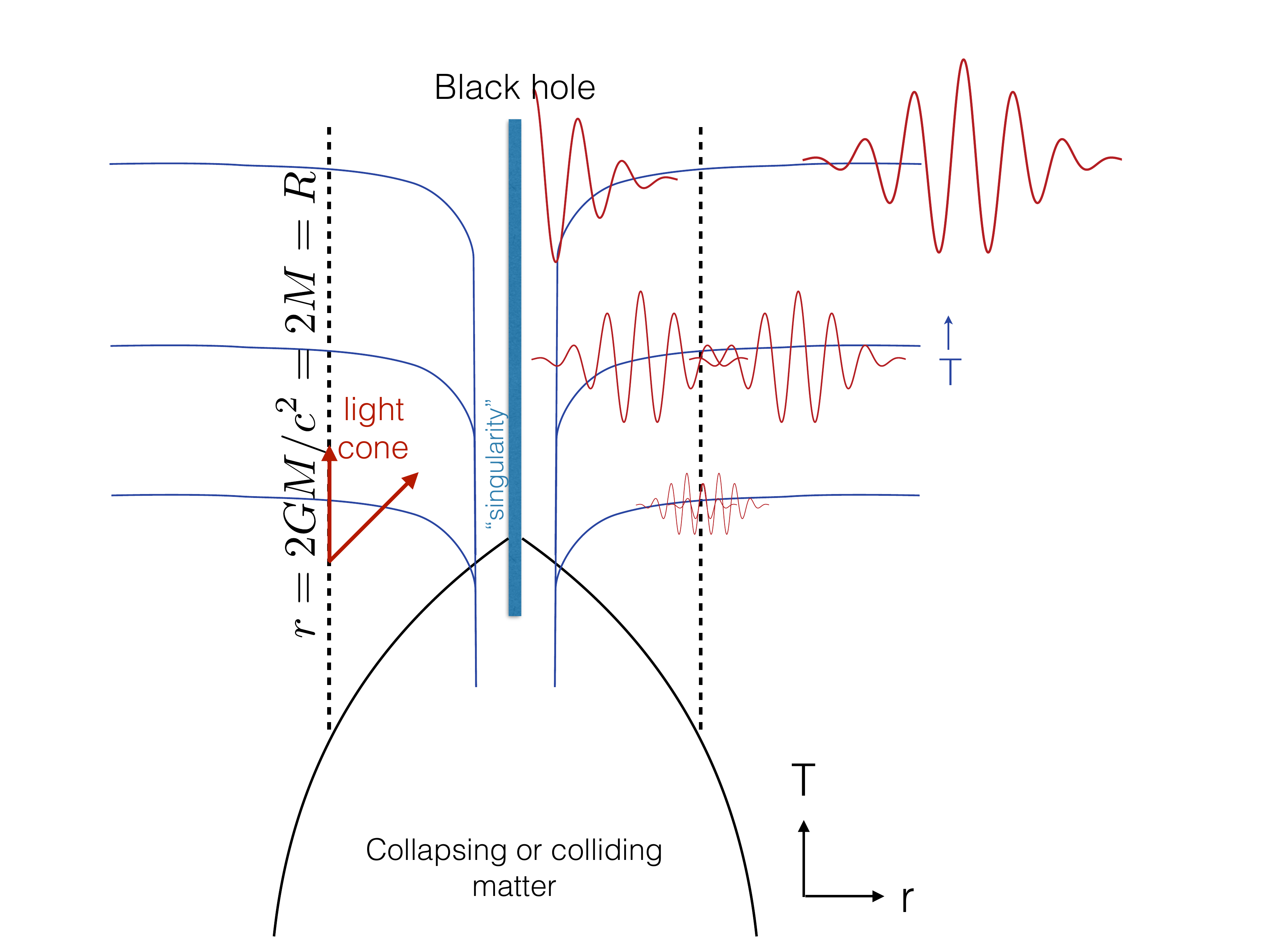}
\caption{A spacetime diagram of a BH, in an Eddington-Finkelstein picture, that forms through matter collapse or collision, and then begins to evaporate.  Shown is a choice of slices used to describe evolution; there is considerable flexibility in the choice of such slices. }
\label{fig1}
\end{figure}

Fig.~\ref{fig1} illustrates the basic phenomenon of BH formation and Hawking radiation\cite{Hawkrad}.  Collapsing or colliding matter forms a BH, with Schwarzschild radius $R$.  We can describe evolution by introducing a slicing of the resulting spacetime, as shown.  In LQFT, fluctuations of any of the quantum fields near the horizon get "pulled apart" due to the strong gravitational field, resulting in outgoing particles together with ingoing excitations that evolve towards $r=0$.  The typical energy of the outgoing particles is given by the Hawking temperature, $E\sim T_H\sim 1/R$.  A "toy" version of the resulting state of such a pair is
\begin{align}\label{Hawkpr}
|\psi\rangle \sim |{\hat 0}\rangle |0\rangle +|{\hat 1}\rangle |1\rangle
\end{align}
where hatted/unhatted kets denote the state of internal/external excitations and we focus only on one possible mode with occupation number zero or one.\footnote{For a more complete description of such a state, see {\it e.g.}, \cite{GiNe}.}  The state \eqref{Hawkpr} is just like the Bell state of an EPR pair, and in particular exhibits the key feature of {\it entanglement}.

Of order one Hawking quantum is emitted per time $R$, building up a state of the form
\begin{align}\label{Hawkst}
|\Psi\rangle_{\rm Hawking} \sim (|{\hat 0}\rangle |0\rangle +|{\hat 1}\rangle |1\rangle)^{\otimes n}\ .
\end{align}
Once $n\sim RM \sim GM^2$ such quanta have been emitted, all the initial energy has been carried away and the BH is expected to disappear.  If it does, the internal excitations are no longer part of the physical description, and the quantum state is found by tracing over them to find a density matrix,
\begin{align}\label{Hawkrho}
\rho_{\rm Hawking} = {\rm Tr}_{\rm bh}\left(|\Psi\rangle_{\rm Hawking}\langle\Psi|_{\rm Hawking}\right)\sim \left( |0\rangle\langle0| +|1\rangle\langle1|\right)^{\otimes n}\ .
\end{align}
However, if this is the fundamental evolution, from what could be an initial pure state of the matter forming the BH to a mixed state, that violates the quantum-mechanical principle of unitarity.

This violation of unitarity can be quantified by the von Neumann entropy of the density matrix $\rho$ of the radiation,
\begin{align}\label{HawkS}
S_{\rm vN} = -{\rm Tr}(\rho \log \rho) =- \Delta I
\end{align}
where we think of $-\Delta I$ as "missing information."  As the BH emits Hawking particles, this missing information grows uniformly in time, until it reaches a final value $S\sim GM_0^2\sim (M_0/m_{\rm pl})^2$, where $M_0$ is the initial BH mass, and $m_{\rm pl}$ is the Planck mass.  This is enormous for a macroscopic BH.

This failure of unitarity is the origin of the  crisis in fundamental physics which we therefore call the {\it unitarity crisis}.  Various possible resolutions have been considered.  The most obvious and mundane possibility is that black holes don't completely evaporate, but instead leave behind microscopic black hole remnants. However, these remnants would have to have an unbounded number of internal states, to parameterize the missing information from an arbitrarily large black hole.  This kind of spectrum leads to a disaster -- unbounded production of such remnants in general physical processes, and is thus apparently ruled out.\footnote{For further discussion, see \cite{WABHIP,Susstrouble}, and references therein.}  

A second possibility is that there is some error in reasoning in the argument that has been just outlined for information loss.  However, over forty years of careful examination have not revealed such an error, and so that seems extremely unlikely.  (Further discussion of one newer argument for an error, the "soft hair" proposal\cite{Hawksoft,HPS1,HPS2,HHPS}, will appear below.)  

The third possibility now seems by far the most likely:  this crisis points to an error in underlying principles, and thus calls for modification of one or more of the principles of relativity, quantum mechanics, and locality.     Since these principles are the cornerstones of LQFT, our current best description of the rest of physics, 
this is clearly the most exciting possibility. It indicates that BHs should help serve as guides to a more profound understanding of physics.  Indeed, it appears that the unitarity crisis may play a role analogous to the crisis of atomic instability in classical physics; there the need to explain the atom led to the conceptual revolution of quantum mechanics.

There are additional reasons to view the unitarity crisis as a key problem for quantum gravity.  First, the quantum physics of BH formation and evaporation appears to be the generic high-energy physics of gravity:  if we consider a general scattering process, and increase the CM energy, at large enough values we expect to enter the regime of strong gravity, where classically a BH would form.  
Secondly, as has become clear, and as we will see below, the unitarity crisis appears to require a modification of {\it long distance}, or infrared, physics.  That strongly suggests it is a more profound problem than that of nonrenormalizability, which has historically motivated a lot of the work in quantum gravity.  Nonrenormalizability was in particular a strong motivator for supergravity and superstring theory, since those theories promised to greatly improve, or cure, the divergences in the perturbative expansion.  But, nonrenormalizability represents a short-distance problem, anticipated to be cured by short distance effects, {\it e.g.} at the Planck or string length scales.  Moreover, when viewed in the perturbative scattering context, the problem of nonunitarity arises from properties of the {\it sum} over diagrams\footnote{For further discussion, see {\it e.g.} \cite{SGErice}, and references therein.}, and in that sense in addition to being a long-distance problem, involves the nonperturbative physics of gravity.

A first possibility for a resolution arising from an error of principles was suggested by Hawking\cite{Hawkunc}, who initially argued for a modification of quantum mechanics, allowing fundamentally nonunitary evolution.  However, it was later argued that such a breakdown of unitarity would lead to massive violation of energy conservation\cite{BPS}, in violent disagreement with experience.

Attempts to modify quantum mechanics\footnote{Beyond \cite{Hawkunc}, see also for example \cite{NLQM}.} have typically led to disaster; quantum mechanics is a remarkably rigid framework.  Because of this "quantum rigidity," and for other reasons such as its many experimental tests, I will therefore make the (radically conservative) assumption that the Universe is governed by quantum-mechanical principles
(suitably generally formulated to incorporate gravity\cite{UQM}).  This means we should look for a different error of principles.

Indeed, let's turn things around.  Suppose that BHs can be thought of as quantum subsystems (this will be discussed further below).  Suppose that, as Hawking argued, they also build up information content, or more precisely entanglement with their environments, and they radiate and disappear.  And, suppose that quantum mechanics -- specifically unitarity -- holds.  This collection of statements implies that information  {\it must} transfer out of a BH as it decays.  This violates the locality principle, which states that information doesn't propagate outside the light cone, and does so not just on microscopic scales but at scales the size of an arbitrarily large black hole.  Thus black holes strongly point to a long-distance modification of conventional locality, which also turns out  to  be indicated by other behavior of gravity.

However, if this is true, a basic principle of LQFT is not valid, and we need to think about how to consistently describe such physics.  Another important question is how to reconcile this statement with the {\it approximate} validity of LQFT on  semiclassical background geometry.

\section{Quantum-first gravity}

To address these problems, one can try to develop a principled approach, based on the assumption that the fundamental description of the theory should be quantum-mechanical.  This is what I call the "quantum-first" approach to gravity\cite{UQM,BHQIUE,SGalg,CCM,CaCa,CaSi,QFG,QGQFA}:  {\it instead of trying to quantize geometry, we should ask what kind of quantum structure can approximately describe gravity.}

Basic ingredients of a quantum theory are a linear space of states, $\cal H$, an algebra of observables, $\cal A$, and, for states with appropriate asymptotics, such as Minkowskian, unitarity.\footnote{For further discussion see \cite{UQM}.}  However, clearly more mathematical structure than this is needed to describe physics.  

\subsection{Localization and subsystems in quantum gravity}

LQFT furnishes an example of the role of such mathematical structure.   As described in the algebraic approach (see {\it e.g.} \cite{Haag}), this mathematical structure arises from the manifold $M$ on which the LQFT is defined.  Associated to any open set $U\subset M$, there is a subalgebra ${\cal A}_U\subset \cal A$ of the observables with support only in $U$ ({\it e.g.} fields smeared with compact-support test functions).  For spacelike separates open sets $U$, $U'$, the subalgebras ${\cal A}_U$ and ${\cal A}_{U'}$ commute.  Moreover, these subalgebras have nesting and intersection properties just like the open sets.  So, the subalgebras "mirror" the topological structure of the manifold given by the open sets, along with the causal structure.  This "net" of subalgebras is the mathematical structure relevant to describing LQFT.

However, this does not appear to be the correct mathematical structure for describing a gravitational theory.  In fact, already at the perturbative level we discover that this local algebraic structure must fail.  Consider a scalar field $\phi(x)$, coupled to gravity.  While $[\phi(x),\phi(y)]=0$ for spacelike $x-y$, $\phi(x)$ is not a gauge invariant observable.  Gauge observables can be constructed\cite{DoGi1},\footnote{For related previous constructions, see \cite{Heem,KaLigrav}.} by "dressing" $\phi(x)$ to give a new operator $\Phi(x)$ that also creates the gravitational field associated to a particle created by $\phi(x)$.  

Since $\Phi(x)$ creates a gravitational field, which extends to infinity, now generically\cite{DoGi1} $[\Phi(x),\Phi(y)]\neq0$ for spacelike $x-y$.  In fact the nonvanishing commutators match well to the "locality bound" that was proposed in \cite{GiLia,GiLib,LQGST}, to characterize the regime where LQFT fails to be an accurate description of physics.

This shows that the fundamental gauge-invariant observables in gravity don't obey a local algebra, and thus we must have some {\it other} mathematical structure on $\cal H$, providing the foundations of the theory.  This "substrate" is not clearly based on any underling manifold structure, despite what is commonly assumed.  In LQFT, the net structure on $\cal A$ is directly associated to the manifold, but for gravity it is an important question what replaces this.

Indeed, without some version of locality, providing a notion of separability, it is not clear how to even proceed to describe physics.  In quantum systems, a more primitive notion of locality is that of {\it subsystems}.  For finite quantum systems, subsystems are described in terms of tensor factorizations of the Hilbert space, such as ${\cal H}={\cal H}_1\otimes{\cal H}_2$.  In LQFT, subsystems can be described in terms of the commuting subalgebras ${\cal A}_U$.

We can ask if there is some suitable definition of subsystems for gravity.  Full discussion of this would take us too far afield; recent discussion of this subject includes \cite{DoGi3,QFG,QGQFA,DoGi4,SGsplit}, which can be briefly summarized.  Consider a state with some matter excitations in a neighborhood $U$, gotten by acting with the field operators creating these matter excitations.  This state must be dressed, as described above, and so if this state is created from the vacuum, the dressing will also create a gravitational field in the complement $\bar U$ of the neighborhood.  This suggests that information about the matter state in $U$ is accessible, by measuring this gravitational field.  However, the preceding references show that, working perturbatively in Newton's constant, one may choose the dressing outside $U$ to just depend on the {\it Poincar\'e charges} of the matter inside $U$.  An analogous statement arises in electrodynamics, where a charge distribution inside $U$ may be dressed with a field configuration outside $U$ that just depends on the total charge $Q$ of the distribution.  This may be shown, for example, by running all the field lines to a common point inside $U$, and then outward in some standard configuration independent of the charge distribution.

This provides an outline of an argument for why the soft charge  proposal\cite{Hawksoft,HPS1,HPS2,HHPS} won't help with the unitarity crisis.  If the gravitational field outside a neighborhood can be taken to be independent of the detailed matter configuration inside the neighborhood, then the soft charges will be likewise independent.  This indicates information can be localized in the neighborhood, at least perturbatively, despite nonlocal behavior of gravity; we could just as well take such a neighborhood to be inside a BH.\footnote{Modulo details of constructing dressings in non-trivial backgrounds.}  If so, the soft charges don't necessarily encode information about the matter inside a BH.  If, on the other hand, they had been found to always encode or exhibit the information inside a BH, then that would have been an argument that standard gravitational physics, studied more carefully, contained a possible resolution to the crisis -- {\it i.e} this would have indicated a resolution via identifying an error of reasoning.

Extending this approach to the general problem of  subsystems in gravity, an initial proposal for the corresponding mathematical structure has been discussed in \cite{QFG,QGQFA,SGsplit}, involving a network of Hilbert space inclusions, ${\cal H}_u\otimes {\cal H}_{u'}\hookrightarrow {\cal H}$, where $u$, $u'$ are nontrivial indices including information about Poincar\'e charges.  This would replace the network of subalgebras of LQFT.  In such a structure, a continuum manifold doesn't necessarily play a fundamental role, and may only arise in the weak-gravity, or "correspondence," limit.  This also raises a question about the role of topology, although conceivably this mathematical structure also has pertinent topological structure.

Clearly this discussion starts to become fairly abstract, and elaboration would lead us further afield, but there are four takeaway messages.  The first is that locality is not necessarily sacred in quantum gravity -- it's not even obvious how to define it.  The second is that nonetheless there appears to be an approximate notion of subsystems in gravity, and that a BH in particular can behave as a subsystem.  Third, we expect there to be  some new mathematical structure providing the foundation for quantum gravity, which is plausibly related to this subsystem structure, and it should match onto the structure of LQFT on semiclassical geometry in the weak-gravity, correspondence, limit.  Finally, some consistent description of black hole like objects  should fit into such a framework; indeed, as suggested above, the need to describe them is expected to give key hints for the structure of the theory!

\subsection{Some postulates for quantum gravity}

Before returning to black holes, it is useful to distill a set of principles on which to proceed, from the preceding discussion.  These can be stated in the form of three postulates for quantum gravity (a fourth will appear in the next section).  For a more careful statement of these see \cite{NVU}.

Postulate I is that quantum gravity respects the principles of quantum mechanics.  The essential ones were outlined above; for further discussion see \cite{UQM}.

Postulate II is that there is, at least approximately, a mathematical structure of subsystems on the Hilbert space $\cal H$.  As indicated, a careful definition of this structure is a little subtle, but we will use a suitable simplification of it below.

Postulate III is the correspondence postulate, which we have also noted:  in weak gravity limits, the structure of the more fundamental theory should approximately reproduce the behavior of LQFT plus semiclassical geometry.  We will seek a minimal departure from the latter behavior. 

A fourth postulate will be introduced below, in describing BH evolution.

\section{Quantum black hole evolution}

We now return to the problem of BHs, to which we will apply the preceding postulates.

Postulate II, Subsystems, is taken to in particular imply that a BH and its environment can be thought of, at least approximately, as subsystems of a larger quantum system.  As we have outlined, the definition of such subsystems in gravity is somewhat subtle.  But, as a simple model, we will assume that the quantum states can be written in the form
\begin{align}\label{BHstates}
|K; M; \psi_e, T\rangle\ .
\end{align}
Here we consider a Schr\"odinger picture description, where states evolve in time $T$; in the geometrical picture, such a description can be given by introducing a time slicing as illustrated in fig.~\ref{fig1}.  The label $K$ describes the internal states of the BH; we expect there to be an enormous number of them, corresponding to the BH entropy, with mass $\leq M$:
\begin{align}\label{densstates}
N = e^{S_{bh}} \ ;
\end{align}
 most of these are expected to be in a range of masses $(M, M-\Delta M)$ with $\Delta M\sim 1/R$.  
The state of the environment is described by $\psi_e$.  Moreover, by Postulate III, Correspondence, this state should have a good approximate description by a state of LQFT.

Postulate I, Quantum Mechanics, implies in particular that evolution of states \eqref{BHstates} is unitary.  Its infinitesimal generator takes the form
\begin{align}\label{Hdef}
H=H_{bh} + H_{env} + H_I\ ,
\end{align}
where $H_{bh}$ and $H_{env}$ act on the BH and environment subsystems, respectively, and $H_I$ acts on both, and can transfer information between them.  In fact, LQFT evolution, on a time slicing, can be put in this form\cite{NVU}.  However, LQFT evolution will not be unitary once the shrinkage of the BH is taken into account, as we have argued above.  In LQFT, $H_I$ only increases entanglement between BH and environment, either by creating Hawking pairs, or by transferring entanglement from the environment to the BH, resulting in the unacceptable situation described in section \ref{sec2}.  So, in order to obey Postulate I, \eqref{Hdef} must have another term that transfers entanglement from the BH to the environment: BH quantum states must be able to influence their surroundings.  Moreover, the LQFT version of $H_{bh}$ is not expected to give the correct dynamics, although we will be largely agnostic about its exact form.  Finally, in LQFT, the state on a slice in principle allows a description of a much larger number of BH states than given by \eqref{densstates}; in assuming that the states are described as in \eqref{BHstates} and \eqref{densstates}, we are also assuming an important structural departure from the LQFT description.

We will parameterize the required information-transferring interactions, working in an effective picture, which describes perturbations of the background BH.  In effect, we can think of the BH in analogy to a big atom, albeit one with a very dense spectrum of states, and parameterize its couplings to its environment.

The simplest kind of interaction that can transfer information from BH to environment is a product of operators acting on the respective subsystems.  Let $\lambda^A$ be the $U(N)$ generators, which give a basis for operators acting on BH states; these can be coupled to operators $\calo^b(x)$ acting on the environment, which are well-approximated as operators of LQFT.  We will adopt a strategy of "parameterizing our ignorance" by introducing an additional term in $H_I$ that is a general superposition of such bilinear operators:
\begin{align}\label{DHdef}
\Delta H_I = \sum_{Ab} \int dV \lambda^A \calo^b(x) G_{Ab}(x)\ ,
\end{align}
where the integral is over a time slice in the environment; $dV$ is the background volume element.  At this stage, the coefficients $G_{Ab}(x)$ are arbitrary; they act as "structure functions" of the quantum BH states.  However, we will constrain these functions by further consideration of the postulates.

At this point we should also note that, when viewed from the perspective of the semiclassical BH geometry, such as illustrated in fig.~\ref{fig1}, the couplings \eqref{DHdef} are nonlocal, at least on scales comparable to $R$.  This is apparently inevitable, in order to save quantum mechanics.  Of course, this raises a further concern, regarding consistency.  In flat Minkowski space, transfer of information outside the light cone can be converted, via a Lorentz transformation, into transfer of information that is backwards in time; combining such transfers allows, in principle, an observer to send a signal to their backward lightcone, raising causality paradoxes.  However, a key point is that this argument uses the Lorentz symmetry of Minkowski space.  This is not a symmetry of the BH background, on which we are considering propagation, and in this background the argument that nonlocality implies acausality no longer holds.  Thus, it remains quite plausible that such interactions of BH states with their surroundings do not necessarily introduce deep inconsistencies associated with acausality\cite{NLvC}\cite{BHQIUE}.

As a first constraint on the couplings \eqref{DHdef}, we will  introduce one further postulate.  Postulate IV is a statement of universality:  we assume that the new interactions in \eqref{Hdef} couple universally to all matter and gauge fields (including perturbative gravitons).  This seems like a natural postulate given the universal nature of gravity.  However, it can be even more strongly motivated.  

One motivation arises from considering Gedanken experiments, involving "black hole mining"\cite{UnWamine,LaMa,FrFu,Frol}.  The simplest example comes from imagining that we thread a BH with a cosmic string.  In that case, the rate at which the BH loses mass increases, because there are additional modes along the string in which Hawking radiation is produced.  However, if we increase the rate at which the BH decays without increasing the rate at which information transfers from it, we return to the basic problem of trapped information at the end of evaporation, which precipitated the crisis.  This is avoided if the new interactions couple universally to all matter, so introduction of new modes that can carry energy also means that these modes can transfer information.  

A second motivation comes from the desire to preserve, at least approximately, the beautiful story of BH thermodynamics.  If the new interactions coupled only to certain fields, for example photons or gravitons, that presents problems for detailed balance, if one tries to bring a BH in equilibrium with a thermal bath.

The simplest way to implement universality is if the interactions couple to the energy-momentum tensor, and so \eqref{DHdef} is replaced by the specific form
\begin{align}\label{HTcoup}
\Delta H_I =  \int dV\sum_A  \lambda^A G_{A}^{\mu\nu}(x) T_{\mu\nu}(x)\ ,
\end{align}
where the stress tensor $T_{\mu\nu}$ includes perturbative gravitons.  In that case, we find that the operator-valued quantity
\begin{align}\label{Hmndef}
H^{\mu\nu}(x)= \sum_A  \lambda^A G_{A}^{\mu\nu}(x) 
\end{align}
behaves like a metric perturbation that depends on the quantum state of the BH.  The Postulates supply additional conditions on $H_{\mu\nu}(x)$. 

 Postulate III, Correspondence, tells us that the information-transferring couplings $G_{A}^{\mu\nu}(x)$ should be localized near the BH; they shouldn't, for example, describe transfer of information from a BH to the next galaxy, which would be a more extreme departure from LQFT.  We therefore suppose that these couplings are supported only to a radius
\begin{align}\label{Radef}
r=R_a=R+\Delta R_a\ ,
\end{align}
and that $R_a$ is not enormously larger than $R$.  Moreover, if $\Delta R_a$ is finely tuned, {\it e.g.} to a microscopic value $\Delta R_a\ll R$, then the couplings \eqref{DHdef} or \eqref{HTcoup} will produce high-momentum excitations, very near the horizon.  This can be thought of as a way of modeling behavior like that of a firewall\cite{SGTrieste,Braunstein,AMPS}.  In addition to requiring an unexplained fine-tuning of $\Delta R_a$, this also implies a dramatic breakdown of the geometry near what would-be the horizon, and in particular also violates the Correspondence postulate.  So, we will assume that $\Delta R_a$ grows with $R$, {\it e.g.} as a power; the simplest choice is $\Delta R_a \sim R$.  We moreover assume that the couplings   $G_{A}^{\mu\nu}(x)$ don't vary on microscopic scales, and {\it e.g.} only vary on scales $\sim R$.  Finally, also to avoid high-energy excitations near the horizon, and to minimize the departure from LQFT predictions, and particularly from BH thermodynamics, we assume that the couplings dominantly describe transitions between states with energy difference $\Delta E\sim 1/R$.

Postulate I, and specifically unitarity, implies additional conditions.  The couplings \eqref{HTcoup} (or \eqref{DHdef}) are required to transfer information from the BH at a rate that overcomes the Hawking entanglement growth, and so at a rate of size\footnote{Here we can define the BH's information $I$ in terms of its entanglement with the environment; this information must decrease to zero by the time the BH decays.} $dI/dt \sim 1\, {\rm qubit}/R$.  In this sense, the required information transfer is a significant effect, as emphasized by \cite{Pageone,Pagetwo,Mathur}.  

The simplest way to achieve this rate\cite{NVNLT} is if the metric perturbation in a typical BH state,
\begin{align}
\langle H_{\mu\nu}(x)\rangle = \langle\psi, T|H_{\mu\nu}(x)|\psi, T\rangle\ ,
\end{align}
which is dimensionless, is also $\calo(1)$.  This follows essentially from dimensional analysis, if the spatial and temporal variation scales of $H_{\mu\nu}(x)$ are all of order $R$, and is further discussed in \cite{NVNLT}.

However, we can ask if $\langle H_{\mu\nu}(x)\rangle=\calo(1)$ is {\it required} to achieve sufficient information transfer.  Here we encounter a general problem in quantum information theory.  Suppose we have two subsystems $A$ and $B$, with $A$ much smaller than $B$.  Suppose that the two systems evolve by a hamiltonian 
\begin{align}
H=H_A+H_B+H_I\ ,
\end{align}
where $H_A$, $H_B$ act only on their respective subsystems, and $H_I$ couples the two.  Suppose moreover that the dynamics of $H_A$ and $H_B$ are sufficiently random to distribute information efficiently.  Then, how does the information transfer rate from $A$ to $B$ depend on the couplings in $H_I$ and on the energy scales?  Refs.~\cite{NVU} conjectured a rate quadratic in the couplings, which has been checked in simple toy models in \cite{GiRo}.  

For our current BH purposes, we will estimate the rate by noting that \eqref{HTcoup} (or \eqref{DHdef}) cause transitions where the BH emits a quantum. When it does so, we expect that $\calo(1)$ qubit of information can be transferred.  So, the information transfer rate is expected to be approximated by the rate at which transitions occur.  The latter can be estimated by Fermi's Golden Rule, so
\begin{align}\label{rateest}
\frac{dI}{dT}\sim \frac{dP}{dT} = 2\pi \rho(E_f) |\Delta H_I|^2\ ,
\end{align}
where $|\Delta H_I|$ is the typical size of a matrix element of $\Delta H_I$.  We require this rate to be the relatively large value $\sim 1/R$.  However, the density of final states $\rho(E_f)$ in \eqref{rateest} includes a factor of the density of BH states that we transition to, and the latter is expected to take the enormous value $\sim N$ of \eqref{densstates}.  That means that the typical matrix elements of $\Delta H_I$ can be correspondingly {\it small},
\begin{align}
|\Delta H_I| \sim \frac{1}{\sqrt N} = e^{-S_{bh}/2}
\end{align}
-- a tiny value!  Moreover, if  $H_{\mu\nu}$ behaves like a generic operator on the space of BH states, it has a typical expectation value that is correspondingly exponentially small,
\begin{align}\label{Hweak}
\langle H_{\mu\nu}(x)\rangle = \calo (e^{-S_{bh}/2})\ .
\end{align}

In summary, we have found two scenarios for such "quantum halos" around BHs, in which interactions transfer information from BH quantum states to the BHs environment.  The simplest, with $\langle H_{\mu\nu} \rangle\sim 1$ we refer to as the "strong/coherent" scenario; here the perturbation in the effective metric is substantial, and behaves like a classical perturbation to the metric.  The minimal, with $\langle H_{\mu\nu} \rangle\sim \exp\{-S_{bh}/2\}$ we refer to as the "weak/incoherent" scenario.  Both of these scenarios raise some important questions.

The first is how to understand such effects, from a more fundamental perspective.  Here there is more to learn, but the preceding discussion has hinted at a modification of how one thinks about locality and transfer of information, compared to how it is described in a semiclassical background geometry.  Fundamental gravitational physics is likely not based on classical spacetime, which only arises approximately, in a certain limit; taking spacetime too literally may lead to errors in the description, which are accounted for by the effects we are parameterizing.  A deeper understanding likely requires a more intrinsically quantum view of information localization and transfer, and of spacetime itself.  If, as expected, it is true that the fundamental description of physics does not involve spacetime, and instead involves different structure on Hilbert space, it may be that a description of the departures of this structure from a spacetime description cannot be simply described in spacetime terms.  Some initial exploration of these questions appears in \cite{QFG,QGQFA}.  A related question is whether some of these ideas might be actually realized in the AdS/CFT correspondence.

A second question is whether such unitarizing effects, since they apparently need to reach well outside the horizon, might have any observational consequences.

\section{Observational probes}

\subsection{Weak/incoherent scenario}

In the weak/incoherent scenario the effective metric perturbations are tiny, as in \eqref{Hweak}. Since they do not behave like large, classical fluctuations, that suggests that they have negligible effect for astrophysical black holes.  

However, to take a closer look, consider scattering of an excitation, {\it e.g.} a photon, from such a quantum halo.  This scattering probability can also be estimated by Fermi's Golden rule\cite{SGEHT}, and so will be of the form
\begin{align}\label{sprob}
\frac{dP}{dt}=2\pi \rho(E_f) \left| \int dV \langle K| H^{\mu\nu}|\psi\rangle \langle\beta|T_{\mu\nu}|\alpha\rangle\right|^2\ ,
\end{align}
where $|\alpha\rangle$ and $|\beta\rangle$ are the initial and final states of the excitation.  While the matrix element of $H^{\mu\nu}$ is again exponentially small, that is once again compensated  by the enormous density of states.  Thus, this probability can also be of order $1/R$ -- and correspondingly there can be $\calo(1)$ modifications to the scattering cross section.

However, we saw that the Correspondence postulate implied that $H_{\mu\nu}$ has spatial and temporal variation scales $\sim R$, and thus the momentum transfer in such scattering is $\Delta p\sim 1/R$.  Consider the situation for the electromagnetic image of M87.  EHT is observing photons of wavelength $\sim 1\, mm$.  However, the radius $R$ for the central object in M87 is $\sim 2\times 10^{10}\, km$.  So, a momentum transfer of this size is utterly negligible for such photons.

Note, though, that in the case of gravitational radiation from BH mergers, radiation wavelengths comparable to the BH size contribute importantly.  At such wavelengths, this scattering can be a significant effect.  This suggests that the weak/coherent scenario can be probed by investigating the effect of such modifications, {\it e.g.} to absorption cross sections, on observed gravitational wave signals.

\subsection{Strong/coherent scenario}

If the strong/coherent scenario is correct, its effects could be even more pronounced.  This scenario predicts fluctuations in the effective metric, of size $\Delta g_{\mu\nu}\sim 1$, extending a distance $\sim R$ from the horizon, and fluctuating on spatial and temporal scales $\sim R$.  Such classical-like fluctuations can have important effects on light propagation, and in particular on EHT images.  Simulations of the image distortions, with a simple model for such perturbations, indeed reveals the possibility of dramatic effects for sufficiently large perturbations\cite{GiPs}.  A particular observational target is the temporal variation, on time scales given by the typical frequencies of the effective perturbations.

If one assumes these fluctuations should lead to a minimal departure from Hawking's description of BH evolution, and in particular should not greatly alter the thermal spectrum of the BH emission\cite{GiPs}\cite{SGEHT}, the frequencies of the fluctuations should be of comparable to the thermal frequency,
\begin{align}
\omega \sim \omega_T = {1\over 4\pi M}\, {\sqrt{1-a^2}\over 1+\sqrt{1-a^2}}
\end{align}
for spin parameter $a=J/M^2$.  The two main targets of EHT are Sgr A${}^*$ and M87, for which the corresponding periods are 
\begin{align}\label{periods}
\begin{split}
 P= \frac{2\pi}{\omega_{\rm T}}  
 & \simeq 0.93 
\left(\frac{M}{4.3\times 10^6 M_\odot}\right)\left({1\over 2} + {1\over 2\sqrt{1-a^2}}\right)~{\rm hr}\cr
 &\simeq  59 \left(\frac{M}{6.5\times 10^9 M_\odot}\right)\left({1\over 2} + {1\over 2\sqrt{1-a^2}}\right)~ {\rm d}
 \end{split}
\end{align}
The period for Sgr A${}^*$ is less than the EHT image scan time of several hours, making it hard to resolve such temporal fluctuation.  This suggests the best sensitivity to such variation is from M87 observations.

The recently released\cite{EHT} EHT image spans a period of seven days, which is short as compared to the M87 period \eqref{periods}; we therefore don't expect it to strongly bound such fluctuations, although some modest bounds appear to be possible given that the seven-day images are close to what is predicted from classical GR\cite{SGEHT}.  Clearly longer duration observations will be of interest to increase such sensitivity.

\subsection{General comments}

While the preceding discussion has sought to provide strong motivation for a picture of BH evolution, it is worth considering the situation even more broadly.  The past few years have seen the opening of {\it two} observational windows on a regime for which there were previously none:  the strong gravity regime, near what is classically expected to be a BH horizon.  Over a slightly longer time, the theoretical community has also reached a widespread (though still not universal) view that to reconcile BHs with quantum mechanics, modifications of a description via LQFT on semiclassical geometry must reach out to scales comparable to the horizon size.  While the preceding effective description of quantum dynamics seems perfectly reasonable, any other alternative should also extend into the region near the horizon.  Thus, it is worth actively investigating possible observational constraints on any such scenario, if it is sufficiently well-formulated to start to make predictions.

\section{Conclusion}

The problem of quantum BH evolution suggests that there is something deeply wrong with a semiclassical, geometrical description of BHs, and that BHs are likely intrinsically quantum objects {\it at horizon scales}.  Indeed, in an inherently quantum-mechanical approach to physics, spacetime may not play a fundamental role, and may be only approximately correct, and the strong gravity regime of a BH may begin to particularly reveal this problem.  If we adopt quantum mechanics as a postulate, and demand unitary evolution, we seem to be directly led to a picture where a BH must have interactions with its environment that are outside the standard description of BHs.  Interestingly, the necessary new couplings can be much smaller than na\"\i vely expected, and still save quantum mechanics.  Whether large or small, they potentially have observational consequences, and it is of great interest to investigate these further.\vskip6pt

\enlargethispage{20pt}

%
%

\competing{The author(s) declare that they have no competing interests.}

\funding{This material is based in part upon work supported in part by the U.S. Department of Energy, Office of Science, under Award Number {DE-SC}0011702.  }

\ack{I thank  the CERN theory group, where this work was carried out, for its hospitality.}

%
\bibliographystyle{utphys}
\bibliography{bhqu}

\end{document}